\documentclass[conference]{IEEEtran}
\ifCLASSINFOpdf
  \usepackage[pdftex]{graphicx}
  \graphicspath{{./}}
  \DeclareGraphicsExtensions{.pdf,.jpeg,.png}
\else
\fi
\usepackage[table]{xcolor}

\usepackage{multirow}
\usepackage[caption=false,font=footnotesize]{subfig}
\usepackage{colortbl}

\begin{document}
%

\title{On the Feasibility of CubeSats Application Sandboxing for Space Missions}

\author{\IEEEauthorblockN{Gabriele Marra\IEEEauthorrefmark{1},
Ulysse Planta\IEEEauthorrefmark{1},
Philipp Wüstenberg\IEEEauthorrefmark{2} and 
Ali Abbasi\IEEEauthorrefmark{1}}
\IEEEauthorblockA{\IEEEauthorrefmark{1}CISPA Helmholtz Center for Information Security}
\IEEEauthorblockA{\IEEEauthorrefmark{2}Chair of Space Technology, Technische Universität Berlin}}


\IEEEoverridecommandlockouts
\makeatletter\def\@IEEEpubidpullup{6.5\baselineskip}\makeatother
\IEEEpubid{\parbox{\columnwidth}{
    {\fontsize{7.5}{7.5}\selectfont Workshop on Security of Space and Satellite Systems (SpaceSec) 2024 \\
    1 March 2024, San Diego, CA, USA \\
    ISBN 979-8-9894372-1-4 \\
    https://dx.doi.org/10.14722/spacesec.2024.23033 \\
    www.ndss-symposium.org}
}
\hspace{\columnsep}\makebox[\columnwidth]{}}

\maketitle

\begin{abstract}
This paper details our journey in designing and selecting a suitable application sandboxing mechanism for a satellite under development, with a focus on small satellites. Central to our study is the development of selection criteria for sandboxing and assessing its appropriateness for our satellite payload. We also test our approach on two already operational satellites, Suchai and SALSAT, to validate its effectiveness. These experiments highlight the practicality and efficiency of our chosen sandboxing method for real-world space systems. Our results provide insights and highlight the challenges involved in integrating application sandboxing in the space sector.

\end{abstract}


%

\section{Introduction}
\label{sec:intro:mission_description}

Satellites, now numbering over 10,000 as of 2024~\cite{numberOfSatellites}, have transitioned from extraordinary space achievements to common orbital fixtures, especially with the surge in small satellites like CubeSats and nanosatellites. This accessibility has allowed diverse entities, including universities and startups, to engage in space projects. However, the ease of developing these smaller satellites often comes at the cost of security, making them prone to cyberattacks. Teams behind these projects may lack comprehensive cybersecurity knowledge, leading to significant vulnerabilities. Furthermore, the evolving nature of cybersecurity means satellite software can quickly become outdated, with updates in orbit posing a challenge, as noted in research like Willbold et al.~\cite{SpaceOdyssey}.

Concurrently, there has been a significant evolution in satellite on-board computing, particularly in processing power. This advancement enables small satellites to run full operating systems like Linux, a shift from the basic systems in earlier models. This technological progress enhances satellite functionalities but also adds complexity and vulnerability, necessitating stronger security measures. As systems become more sophisticated, they are more susceptible to threats, requiring a layered defense approach. Sandboxing is one of the effective methods to isolate software vulnerabilities and protect these advanced systems.

In this paper, we discuss the process of selecting a sandboxing mechanism for a satellite project currently under development, named RACCOON \cite{tu_berlin_raccoon}. The project's goal is to design and develop a payload system specifically for small satellites, enabling the global transmission of security keys through a secure and robust communication system. The objective is to create and evaluate a secure and compact satellite. This demonstrator will feature a cognitive radio connection that is resilient to jamming and other forms of interference. This satellite will globally distribute encryption keys, securing communication with remote users against quantum computer-based decryption attacks, particularly catering to operators of critical infrastructure and companies with highly sensitive data. 
The payload of the satellite incorporates an embedded computer system that operates on a Linux distribution and will run on an ARM processor. The mission's software architecture is modular, utilizing the Robot Operating System 2 (ROS 2) as its core middleware~\cite{ros2}. This facilitates efficient communication among various application-specific modules within the system.
There are six planned applications for the payload that include a CCSDS interpreter, Cognitive Radio Control, Satellite Bus Interface, Housekeeper, Key Distribution Algorithm, and a Master Node. Given the sensitivity of transmitted data, the Key Distribution Algorithm and Master Node must be rigorously protected to prevent potential exploitation of vulnerabilities in other applications, particularly if the authentication method is compromised. 
In this paper, we outline a basic set of requirements for application sandboxing within our payload. By defining these requirements, we are able to identify and select a suitable sandboxing mechanism that is appropriate for our payload's security needs. Since the satellite payload implementation is not yet finalized, we tested our sandboxing environment on two specific satellites: Suchai~\cite{suchai_mission} and SALSAT~\cite{salsat_mission}. Suchai, is Chile's first CubeSat developed by the University of Chile (a 1U CubeSat). SALSAT, is a 12 kg satellite from the Technical University of Berlin that carries the spectrum analyzer SALSA for global frequency spectrum analysis, generating heatmaps of spectrum usage and detecting interference. 
These two satellites share a middleware-like architecture design similar to our currently planned satellite payload, providing a relevant context for testing and evaluating the sandboxing mechanism. 
Our evaluation shows that using our chosen sandboxing mechanism could prevent an attacker from getting complete control of the CubeSat, even in case of a successful exploit of a code vulnerability. We also explore different strategies for integrating this mechanism into our current mission and SUCHAI and SALSAT frameworks.

\section{Technical Background}
\label{sec:background}

\subsection{Sandboxing}
In cybersecurity, sandboxing concept emerges as a critical mechanism for safeguarding computing systems. This concept operates on the principle of containment, where software activities are confined to the sandbox, preventing any potential damage to the broader system. Practical implementations of sandboxing can be observed in everyday technologies such as web browsers, which are used to run scripts from untrusted websites safely.

Within the Operating System (OS) environment, sandboxing concept manifests in three forms: complete virtual machines, containers, and application sandboxes. 
 \begin{itemize}
\item Virtual Machines (VMs) and Hypervisors: VMs simulate entire operating systems, providing the highest isolation level among the three methods. While they offer robust security, their significant computational overhead can be a drawback and usually require specific hardware support and acceleration (like Intel VT-x or AMD-V)~\cite{neiger2006intel, fisher2006hardware}.
\item Containers (e.g., Docker, Podman): Containers encapsulate an application and its dependencies in a separate environment while sharing the host system's kernel. Although containers are more secure than running applications directly on the host \cite{DBLP:journals/corr/abs-1804-05039}, they are generally considered less safe than virtual machines or dedicated application sandboxes \cite{DBLP:journals/corr/Bui15, wan2019practical}. This is due to their shared kernel architecture, which can be a vector for vulnerabilities that affect both the container and the host.

\item Application Sandboxes: Application sandboxes create tightly controlled environments to run individual applications, leveraging various OS features to confine the application and restrict its access to system resources. These are also one of the most resource-efficient options, making them particularly suitable for embedded systems with limited computational resources~\cite{vokorokos2015application}.
\end{itemize}

This paper will primarily focus on application sandboxing within Linux OS as it is relevant to our satellite mission.
Using a hypervisor would introduce excessive computational overhead and restrict our choice for the System on a Chip (SoC). At the same time, containers cannot ensure the necessary level of security.
For future reference, anytime we use the term satellite in this paper, we are referring to CubeSat satellites as our satellite is falling into this category. 

\subsection{Application Sanboxing in Linux}
\label{linux_kernel_features}
Application sandboxing at the system level utilizes a range of technologies to achieve strong isolation and security. Many of these technologies are core components of the Linux kernel, forming the foundation for various sandboxing tools to develop their features. Commonly, the kernel features employed by application sandboxes include:

\begin{itemize}
    \item \textbf{Linux Namespaces:} Namespaces are foundational to process isolation in Linux, providing a layer that segregates system resources such as network interfaces, process IDs, mount points, and inter-process communication. The available namespaces include Mount, PID, Network, IPC, User, UTS, and Cgroup namespaces~\cite{sun2018security, raknes2016nsroot}.
    
    \item \textbf{Filesystem Constraints and chroot:} These constraints prevent unauthorized access to critical system files by allowing access only to specific regions of the filesystem, limiting writing permissions to read-only areas. The additional use of \textit{chroot} command changes the apparent root directory for a running process, further enhancing the isolation of sandboxed applications~\cite{docker20164}.
    
    \item \textbf{Resource Limits with cgroups:} The use of cgroups (Control Groups) in setting resource limits, such as CPU and memory usage, is vital in mitigating resource abuse (e.g., Denial of Service)~\cite{rosen2013resource}.
    
    \item \textbf{seccomp-bpf Filters:} The seccomp-bpf (Secure Computing with Berkeley Packet Filters) is a security mechanism that restricts the system calls available to an application~\cite{jia2023programmable}.
    
    \item \textbf{Linux Capabilities:} Linux capabilities divide the privileges of the root user into distinct units, allowing processes to perform specific privileged actions without needing full root privileges~\cite{linux_capabilities}. 
    
    \item \textbf{Network Isolation and Restrictions:} Network namespaces provide a means of isolating the network stack, including IP addresses, routing tables, and interfaces. Network interfaces could then be cloned or moved into other namespaces with the use of virtual network interfaces (MACVLAN).
\end{itemize}

\begin{figure}[htbp]
  \centering
  \includegraphics[width=7.5cm]
  {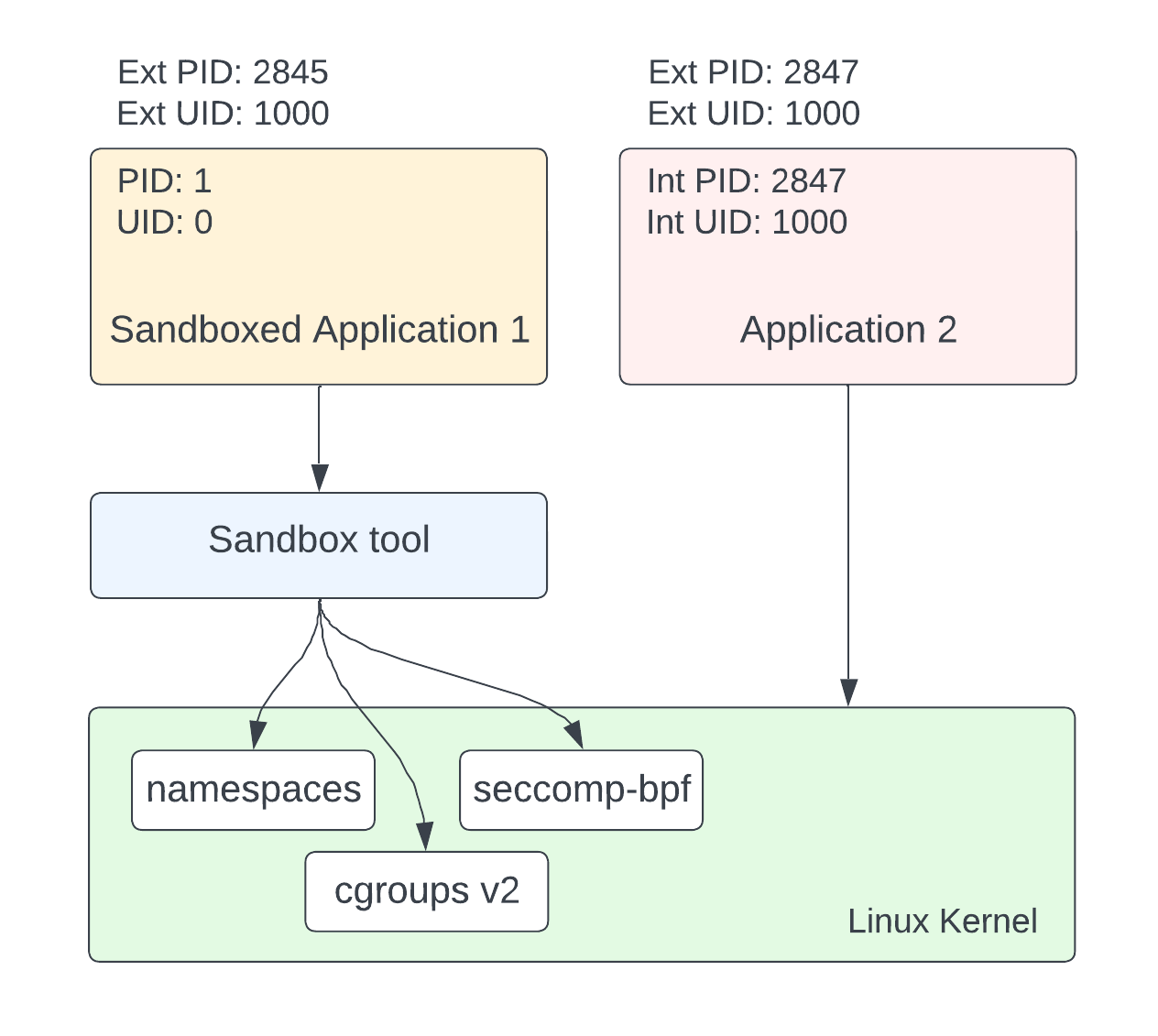}
  \caption{Example of the stack of an application sandbox}
  \label{fig:nsjail_diagram}
\end{figure}

By integrating all, or a subset, of these Linux features and commands, sandboxing tools are able to craft comprehensive sandbox environments. In Figure~\ref{fig:nsjail_diagram}, the diagram illustrates the operational distinction between an example sandboxed application and a non-sandboxed one. Within the sandbox, the application perceives altered Process IDs (PIDs) and User IDs (UIDs), which are essential for isolation (by using Linux kernel Namespaces). PIDs uniquely identify each process, while UIDs designate user privileges. For instance, the sandboxed application operates under the impression of having root privileges (UID 0). However, the actual execution takes place with standard user privileges (UID 1000), with no real root-level access to the host system.
The example sandbox employs Linux Namespaces, seccomp-bpf filters, and cgroups to create an isolated environment and limit the access to the Linux Kernel and system resources. Conversely, Application 2, not being sandboxed, interacts directly with the kernel using the system's true PIDs and UIDs. 

While setting up these tools involves custom and detailed configuration for each application — a process that can be time-consuming and complex — the payoff is noteworthy. Once well-configured, these sandboxes enable applications to perform their designated functions efficiently and securely, significantly reducing or eliminating the risk of system-wide exploits deriving from vulnerabilities in the containerized applications~\cite{borate2016sandboxing, vokorokos2015application}.

\subsection{Satellite environment}

\subsubsection{Operational Threats in Space}
The operational environment of satellites presents a unique set of challenges that necessitate careful and specialized solutions. Satellites must operate almost autonomously in space due to the critical inaccessibility for physical repair or maintenance post-deployment. Communication with ground stations is limited to specific periods when the satellite is within the visibility window, constraining data transmission and command reception. 
Single-Event Upsets (SEUs) are a critical concern in satellite operations~\cite{campbell1992single}. They occur when cosmic radiation causes a change in a bit of RAM or FLASH memory~\cite{seu_space}, a phenomenon more likely in the space environment, and can lead to critical data corruption. Such minor alterations can have cascading effects, potentially compromising the entire mission.
 Software-based mitigation strategies become vital in smaller satellites, where physical shielding and hardware redundancy may be limited due to size, weight and cost constraints. Therefore, error detection~\cite{seu_software_detection} and correction algorithms are essential for maintaining data integrity. Implementing watchdog timers and periodic system resets can also help in mitigating the impacts of SEUs.
Given these limitations, satellite software must be designed with SEU resilience in mind. This also applies to sandbox tools, given that a wrong configuration in sandbox tools could prevent communication with the satellite or hinder command execution.

\subsubsection{Cybersecurity Threats in Space}
\label{sec:background:cyber_threats_in_space}

As described by~\cite{pavur2022building}, satellite systems can suffer from range of cybersecurity issues. For example, input manipulation and parsing are significant concerns: often originating from untrusted or malformed inputs received via the satellite's radio module, attackers exploit vulnerabilities in signal processing or data parsing software, leading to remote code execution~\cite{SpaceOdyssey,scharnowskicase}.
Data exfiltration and unauthorized access to data storage and transmission systems are severe threats to the confidentiality and integrity of mission data~\cite{peled2023evaluating,falco2021cubesat}. 
Lateral movement and privilege escalation within the satellite network are also critical concerns~\cite{SpaceOdyssey}. An initial breach in one component can lead to the spread of the attack to other interconnected systems. An attacker can gain broader access to the satellite's systems by exploiting existing interconnections or elevated privileges.
Finally, denial of service attacks, including resource exhaustion and flooding~\cite{gecgel2021intermittent}, aim to disrupt satellite operations. These attacks can overload the satellite's systems with excessive use of RAM, CPU time, or COM bandwidth.

\section{requirements for sandboxing in space mission}

Given the variety of threats CubeSats face, as detailed in previous sections, the adoption of application sandboxing stands out as an optimal solution for enhancing overall security.
By isolating critical components, such as the code handling untrusted inputs from the radio module, sandboxing could effectively minimize the risk of widespread system compromise due to localized vulnerabilities. This containment aspect is crucial in an environment where even a minor breach can have far-reaching and disastrous consequences.

\subsection{Our Mission Specific Attacker Model}\label{sec:req:threat_model}
Building upon the foundational security principles outlined in the mission description (Section \ref{sec:intro:mission_description}) and the list of potential cyber threats in space (Section \ref{sec:background:cyber_threats_in_space}), this subsection elaborates on the specific attacker model relevant to our satellite mission, incorporating assumptions about the attacker's capabilities and limitations.\newline
\textbf{Authenticated Communication}:
Only authenticated messages can be forwarded by the CCSDS interface or Bus Interface. Access to any of the ROS2 nodes and apps requires at least low-level privileges. In contrast, access to the Master and Key Distribution Algorithm Node is restricted to a few authenticated users with the highest privileges. More details regarding the architecture could be found in Appendix \ref{sec:appendix:RACCOON_diagram}.\newline
\textbf{Attacker Capabilities and Limitations}:
\begin{itemize}
    \item Unauthenticated or Low Privilege Access Exploitation: We assume that an attacker could exploit vulnerabilities in the external bus interfaces without needing authentication. Alternatively, an attacker might gain access to credentials with low privileges. Such access could be used to exploit vulnerabilities in standard applications or ROS2 nodes.
    \item Exclusion of High-Level Credential Compromise: Our model assumes that the attacker does not possess the highest-level authentication credentials. This is a critical assumption, as it shapes our focus on safeguarding particularly sensitive nodes like the Master Node and the Key Distribution Algorithm Node.
\end{itemize}
\textbf{Vulnerability Vector Focus}: The primary concern in our security framework is the exploitation of latent vulnerabilities within our codebase. Despite rigorous validation and testing, undetected vulnerabilities may persist. Our model assumes these vulnerabilities as the primary attack vectors, with an understanding that the software on the satellite payload is inherently not malicious due to strict controls over code uploads and updates.\newline
\textbf{Exploit Containment Strategy}: The central objective is not the elimination of vulnerabilities per se but restricting the exploit's reach within the broader system in case of a successful exploit. This is especially critical in protecting the mission's most sensitive components from unauthorized access (i.e. the Key Distribution Algorithm Node and the Master Node). The integrity and isolation of these elements take precedence over the operational continuity of the mission.\newline
\textbf{High-Risk Components}: The components at the most significant risk are those interfacing externally: the CCSDS interpreter node and the Satellite Bus Interface Node. Given their exposure, these nodes are considered prime targets for exploitation.\newline
\textbf{Internal Application Security}: Acknowledging that specific applications and nodes may be developed by individuals with varying levels of expertise, we must consider the possibility of severe internal vulnerabilities. Hence, these applications must be blocked from accessing critical nodes as well.

\subsection{Space-Related Requirements and Solutions:}
In the context of sandboxing for satellite security, it is important to first establish an environment resilient enough to support such systems. Critical to this resiliency are considerations like protection against SEUs, which could alter critical signatures or constraints that sandboxes impose on applications. 
To mitigate this risk, the satellite system must be equipped with SEU-resilient components before implementing a sandbox. This includes integrating memory solutions such as ECC (Error-Correcting Code) or FRAM (Ferroelectric RAM)~\cite{gupta2017analysis}, known for their robustness against radiation-induced errors. The filesystem itself also requires attention, emphasizing the incorporation of Forward Error Correction (FEC) capabilities. Some filesystems designed for embedded device flash incorporate FEC mechanisms, such as Yaffs (Yet Another Flash File System), JFFS2 (Journalling Flash File System version 2), ELOFS, or F2FS~\cite{olivier2012benchmarking}.
In addition to the filesystem and memory resilience, implementing a configuration file for the sandbox setup is a wise approach. The use of file configuration not only benefits from the filesystem's protection against SEUs but also simplifies the configuration process for developers. 

\subsection{Security Requirements Based on Threat Model:}
\label{sec:req:securityrequirements}
In response to the threats identified for satellite systems~\cite{SpaceOdyssey}, certain security requirements and features are crucial. We will outline these requirements, aligning them with the previously defined threat model for our mission, with a focus on a Linux-based system.

\begin{enumerate}
    \item \textbf{Filesystem Access Restrictions}: Limiting access to necessary files and folders, possibly with read-only mounting, should prevent unauthorized data manipulation and exfiltration.
    \item \textbf{chroot Environment}: Utilizing a chroot environment will restrict the view of the filesystem from the sandboxed processes, further isolating them and helping prevent unauthorized data manipulation and exfiltration, potential escalations, and lateral movements within the system.
    \item \textbf{Device Access Control}: Restricting access to specific devices ensures that only authorized processes interact with critical satellite components, preventing lateral movement.
    \item \textbf{Network Access Control}: By restricting network access to predefined entities, we can reduce the attack surface, even if it is less relevant on a satellite system.
    \item \textbf{Syscalls Filtering}: Blocking undesired system calls is a crucial requirement, as it directly prevents certain types of remote code execution and privilege escalation attacks.
    \item \textbf{Namespace Segregation}: Isolating processes using namespaces ensures that activities in one sandbox do not interfere with others, safeguarding against lateral movement within the satellite network.
    \item \textbf{Limited Capabilities}: Granting minimal required privileges, rather than full root rights, aligns with the principle of least privilege.
    \item \textbf{cgroups for Resource Management}: Implementing control groups (cgroups) helps limit resource usage, preventing denial of service attacks through resource exhaustion.
    \item \textbf{Support for Configuration Files}: Enabling the use of configuration files ensures streamlined and secure sandbox setup, aiding in the protection against SEUs and simplifying management for developers.
    \item \textbf{Small Binary Size}: Emphasizing a compact binary size ensures efficiency in resource-limited environments, aligning with the constraints of CubeSat systems.
    \item \textbf{Logging Capabilities}: Maintaining logs for filesystem and syscall violations provides essential insights into security breaches or attempted attacks, aiding post-event analysis.
    \item \textbf{Code Complexity, Verbosity, and Integration}: Balancing code complexity and verbosity is crucial for maintainability and system reliability. Equally important is the code's integration capability within existing frameworks. Overly complex or overly simplistic, non-modular code interferes with practical integration for our mission.
\end{enumerate}

\subsection{Overview of Considered Solutions:}
In our evaluation of potential sandboxing solutions for our CubeSat, we considered five key options: nsjail~\cite{nsjail_github}, firejail~\cite{firejail_github}, bubblewrap~\cite{bubblewrap_github}, AppArmor~\cite{gruenbacher2007apparmor}, and SELinux~\cite{smalley2002configuring}.
The first three solutions - nsjail, firejail, and bubblewrap - share similarities in their reliance on Linux features like namespaces, cgroups, and seccomp-bpf for application isolation, as discussed in Section \ref{linux_kernel_features}.
\begin{enumerate}
    \item \textbf{nsjail}: This tool was initially designed for hosting Capture The Flag (CTF) challenges but has evolved to support broader sandboxing applications. Although not an official Google product, it is hosted on Google's GitHub page, and some of its maintainers are Google employees. nsjail stands out for its robustness and security-oriented design, making it a compelling choice for high-security environments.
    \item \textbf{firejail}: A SUID program, firejail focuses on sandboxing common desktop applications and stands out for its ability to run X11 applications. It comes equipped with hundreds of profiles for numerous desktop applications, demonstrating versatility in application-level security.
    \item \textbf{bubblewrap}: Distinguished as a tool for creating sandbox environments and running unprivileged containers. Its primary use is to provide a secure environment for containerized applications.
\end{enumerate}
In contrast to the above application-specific sandboxes, AppArmor and SELinux are both Mandatory Access Control (MAC) systems, relying on the Linux Security Modules (LSM) framework.~\cite{wright2002linux}
\begin{enumerate}
    \setcounter{enumi}{3}
    \item \textbf{AppArmor}: AppArmor distinguishes itself through its simplicity and ease of use. It operates by defining profiles for programs, controlling their capabilities and access to system resources through a path-based approach.
    \item \textbf{SELinux}: Developed by the NSA, SELinux operates through a policy-based and type-enforcement approach. Although it offers a highly granular and powerful security model, its complexity in terms of configuration and management is notably higher than AppArmor.
\end{enumerate}
Other tools briefly considered include nsroot~\cite{raknes2016nsroot}, SandFS ~\cite{sandfs}, and sandals~\cite{sandals2023}. However, these were quickly discarded due to issues related to their readiness, reliability, or lack of ongoing support.

\subsection{Discarding MAC Policies:}
In the search for an effective sandboxing solution for our mission, both MAC policies, specifically AppArmor and SELinux, were ultimately discarded.
Despite their robust security frameworks, both lack essential features necessary for a comprehensive sandbox solution. A shared shortfall is the absence of namespace segregation, cgroups, and syscall filtering.
Focusing on individual limitations, AppArmor does not provide network restriction capabilities. On the other hand, SELinux presents its challenges: it lacks support for using Linux capabilities instead of full sudo privileges.

Theoretically, these deficiencies could have been addressed by integrating additional tools or developing new software to complement the MAC systems. However, such an approach would entail development, security testing, and maintenance efforts. Moreover, leveraging well-supported and tested open-source software was deemed more advantageous. Given these considerations and the availability of alternative solutions that already encompassed most, if not all, of the desired features, the decision was made to explore the other three solutions (nsjail, firejail, and bubblewrap).

\begin{table*}[]
\centering
\definecolor{lightgray}{gray}{0.9}
\renewcommand{\arraystretch}{1.1}
\small
\caption{Comparison between nsjail, firejail, and bubblewrap}
\label{tab:app_sandboxes_comparison}
\rowcolors{1}{}{lightgray}
\begin{tabular}{|llll|}
\hline
\rowcolor[HTML]{C0C0C0} 
Features                                & nsjail             & firejail                        & bwrap           \\ \hline
Filesystem Access restrictions          & Yes                & Yes                             & Yes             \\ \hline
chroot                                  & Yes                & Yes                             & No              \\ \hline
Devices access restriction              & Yes                & Yes                             & Yes             \\ \hline
Network access restriction              & Yes                & Yes                             & Partially       \\ \hline
Syscalls filtering                      & Yes                & Yes                             & Yes             \\ \hline
Unprivileged Execution                  & Yes                & No                              & Yes             \\ \hline
Namespace segregation                   & Yes                & Yes                             & Yes             \\ \hline
Capabilities instead of full privileges & Yes                & Yes                             & Yes             \\ \hline
cgroups resource usage limit            & Yes                & Yes                             & No              \\ \hline
Support for ARM Architecture            & Yes                & Yes                             & Yes              \\ \hline
Support for config files                & Yes                & Yes                             & No              \\ \hline
Binary size (kB)                        & 860                & 1703 (firejail) + 138 (firecfg) & 71              \\
Logs file for filesystem violations     & No                 & Yes                             & No              \\
Logs file for syscall violations        & External           & External                        & No              \\
Code complexity and verbosity           & Intermediately complex       & Complex             & Simple (one file) \\
\hline
\end{tabular}
\end{table*}

\subsection{Comparative Analysis}
To identify the most suitable option, we compared nsjail, firejail, and bubblewrap. We evaluated those tools against the requirements critical for our space mission previously introduced in Section~\ref{sec:req:securityrequirements}.
We gathered data from different sources, primarily from official resources: the sandboxes websites, documentation, and manual pages. The binary size was quantified using the latest versions compiled on a virtual machine.
In assessing the code complexity and verbosity, specific criteria were applied, focusing on modularity and portability. The evaluation considered the structural organization of the code, the extent of modularity, the number of lines in primary files, and the dependency on external libraries. We rated the modularity into three categories complex, intermediately complex and simple. 
This collated information is systematically presented in Table \ref{tab:app_sandboxes_comparison}, offering a reference for the comparative outcomes.

The comparison reveals that all three tools – nsjail, firejail, and bubblewrap (bwrap) – robustly support filesystem access restrictions, device access restriction, syscalls filtering, namespace segregation, and the use of Linux capabilities instead of providing full root privileges.
However, differences emerge in certain vital areas. Chroot functionality is supported by nsjail and firejail but not by bubblewrap. Network access restriction is fully implemented in nsjail and firejail, whereas bubblewrap offers only partial support.
Talking about unprivileged execution, nsjail and bubblewrap do not require root permission to execute and create the jail, while this is not the case for firejail. Significantly, bubblewrap does not support the cgroups resource usage limit, contrasting with full support in the other two tools. Additionally, support for configuration files is absent in bubblewrap, limiting flexibility compared to nsjail and firejail.
The support for ARM architecture is uniformly present across all tools, ensuring broad hardware compatibility.
The binary sizes vary significantly, with nsjail and bubblewrap maintaining a smaller footprint compared to firejail's larger size.
Regarding logs, we discovered that nsjail and firejail support writing logs on files for their own activities. Additionally, firejail provides logs for filesystem violations as well. Both nsjail and firejail require external commands (such as \texttt{journalctl -ek}) to obtain detailed information regarding syscall violations. Notably, bubblewrap does not facilitate log file creation or similar auxiliary functionalities.
The code complexity and verbosity show a striking difference. Nsjail maintains a balance with mid-level complexity, whereas firejail exhibits higher complexity and verbosity, which might impede understanding and maintenance. Bubblewrap's approach, although straightforward with a single file, might be oversimplified.

\subsection{Selection of nsjail:}
Following the comprehensive evaluation of the sandboxing tools, nsjail emerges as the preferred solution. This selection is grounded in nsjail's comprehensive alignment with the outlined security requirements, alongside its specific advantages when compared to firejail and bubblewrap.
Nsjail meets all the essential criteria for our space mission, including unprivileged execution, full network access restriction, and its advanced seccomp-bpf filter based on Kafel BPF code language. Additionally, nsjail's smaller binary size is advantageous for space systems where resource constraints are a primary concern. Another critical factor in favoring nsjail is the substantial support it receives from its maintainers. 

While firejail initially seemed a strong candidate, its primary design for desktop applications and the necessity for root privileges ultimately limited its applicability for space missions. On the other hand, bubblewrap was excluded from selection due to its lack of support for certain key features, such as chroot and configuration file support, deemed essential for the robust operation of satellite systems and resilience to SEUs.

A series of practical tests were performed to substantiate the theoretical analysis of nsjail. These tests involved defining policies for nsjail and then verifying the effectiveness of the constraints. Some checks were performed by writing simple scripts or using standard Linux system tools. A custom-developed C++ application \cite{marra_isolationtester} was developed for more advanced verification to call specific kernel APIs directly. The application included testing for namespace isolation (such as different UID, PID, and network settings), filesystem constraints (including chroot and bind mounts), and capabilities related to specific actions.

\section{Real-world experiments}
In this section, we delve into the practical assessment of nsjail's applicability and effectiveness within space environments. This investigation is critical in determining nsjail's suitability as a sandboxing solution for our new satellite framework, which has been discussed in earlier sections of this paper.

Central to our satellite framework is its middleware architecture~\cite{middleware_architecture}, in particular leveraging the capabilities of ROS 2 (Robot Operating System 2)~\cite{ros2}. Instead of a monolithic application structure, the middleware approach presents several advantages for satellite systems. It facilitates easier integration and updating of applications, maintaining consistency in predefined interfaces. This modular design not only enhances system flexibility but also supports asynchronous communication, a vital feature for handling operations in intermittent connectivity conditions typical in space. 

\begin{figure}[htbp]
  \centering
  \includegraphics[width=7cm]{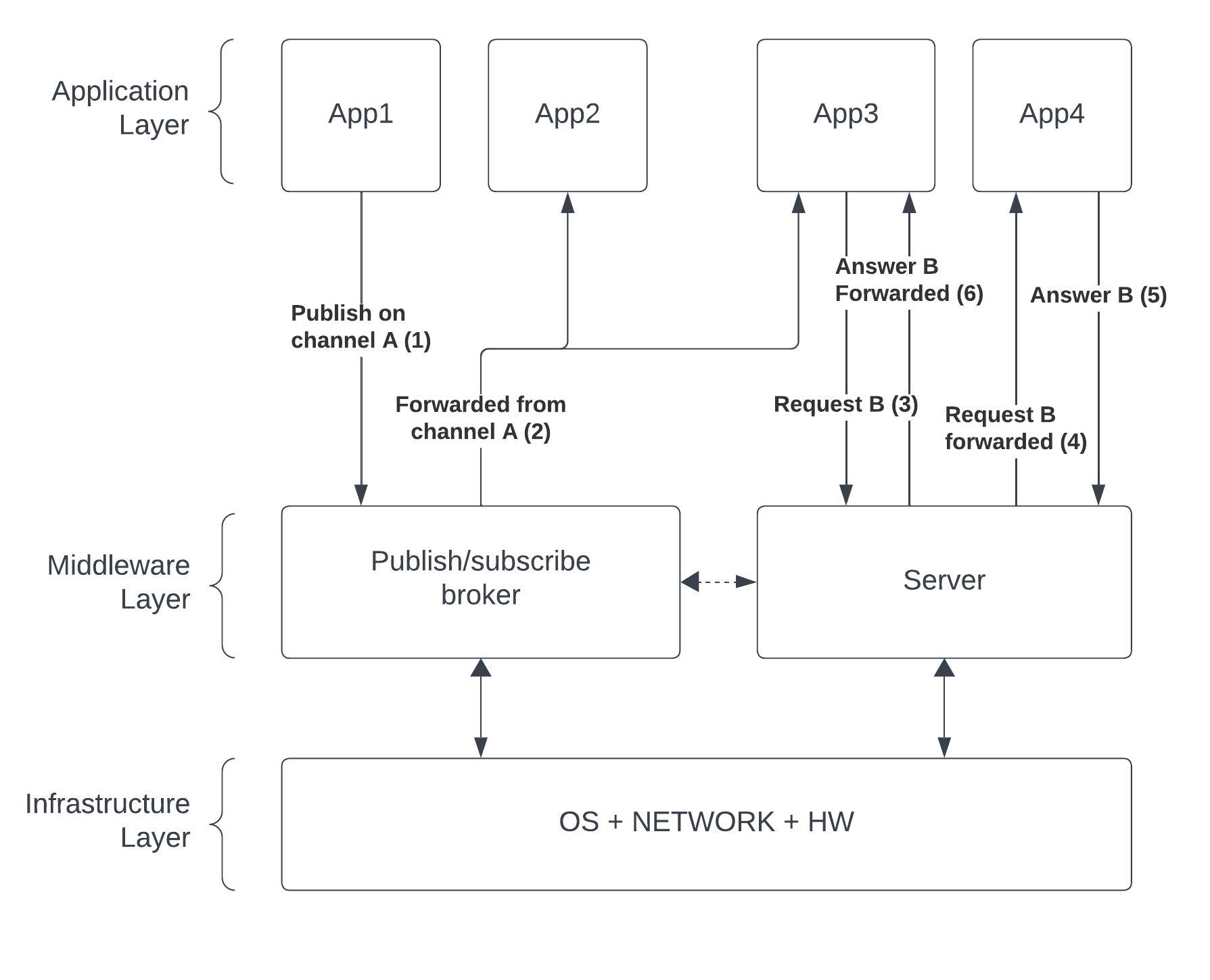}
  \caption{Example of a generic middleware architecture using both publish/subscribe and client/server mechanisms}
  \label{fig:generic_middleware_diagram}
\end{figure}

Figure~\ref{fig:generic_middleware_diagram} demonstrates two of the middleware architecture's communication strategies.
In the first instance, App1, situated in the Application Layer, initiates the publish/subscribe mechanism by broadcasting data on Channel A (Step 1). The Middleware Layer, featuring the Publish/subscribe broker, efficiently relays this data to App2 and App3, both subscribers to Channel A (Step 2), demonstrating the publish/subscribe communication pattern.
Simultaneously, the client/server model is activated as App3, also within the Application Layer, requests Resource B from the Server, a component of the Middleware Layer (Step 3). This request is conveyed to App4 by the Server (Step 4). Following App4's request processing, the response travels back through the Server, which then delivers it to App3 (Steps 5 and 6). This sequence exemplifies the typical request/response flow in client/server interactions.

For example, ROS 2 offers diverse communication mechanisms, such as publish/subscribe (Topics) and client/server models (Services), each catering to different scenarios and use cases.
ROS 2 brings additional benefits to our satellite project: ROS 2 is compatible with multiple programming languages, including C++, Python, and Rust. Furthermore, its compatibility extends across various operating systems, including specialized Linux SoC devices.

Given that our project is still in its development stages and not yet primed for testing, our approach to evaluating nsjail involved leveraging two established frameworks for CubeSats with a similar middleware-like architecture: SUCHAI Flight software and SALSAT IPU software. These frameworks, having been successfully deployed in orbit in 2017 and 2020, respectively, serve as ideal proxies to assess the effectiveness of nsjail in a space environment.
In the upcoming sections, we will present a detailed analysis of the tests conducted with SUCHAI and SALSAT.

\subsection{Experiment with the SUCHAI Flight Software}
The SUCHAI (Satellite of the University of Chile for Aerospace Investigation) project~\cite{suchai_mission} marked Chile's initial venture into CubeSat technology and was successfully launched in June 2017. The SUCHAI Flight software was initially devised for this mission and later became an open-source framework~\cite{suchai_flight_sw}. It provides highly modular command processor architecture, adaptability across platforms like FreeRTOS and Linux, and compatibility with various microcontrollers. The framework combines the CubeSat Space Protocol (CSP) and ZeroMQ (ZMQ)~\cite{hintjens2013zeromq} to provide a middleware like layer, enabling efficient communication among nodes and software components within the CubeSat or on the ground. This setup allows nodes to request data and execute commands. A diagram showing the architecture is available in the Appendix~\ref{sec:appendix:SUCHAI_diagram}.

\subsubsection{Experimentation Setup and Methodology}
In our experiments, the focus was on the Linux port of the SUCHAI software using an Ubuntu 20.04 machine. Our testing procedure began by establishing whether the complete SUCHAI software could be effectively run within an nsjail sandbox. We successfully executed the software in a constrained environment while maintaining it completely functional. The configuration involved the use of chroot, read-only bind mounts, namespaces, and seccomp-bpf.

\subsubsection{Vulnerability Introduction and Impact}
\label{suchai_vulnerability}
To evaluate the effectiveness of nsjail, a deliberate command injection vulnerability was introduced into the command parsing code of the SUCHAI software. This vulnerability simply allowed the execution as code of every text received after a specific sequence of characters, mimicking a possible security flaw. Without a sandbox, an attacker exploiting this flaw could gain shell access and steal the sensitive data from the filesystem.

\subsubsection{Effectiveness of Sandbox Implementation}
A significant improvement in security was noted after integrating the modified SUCHAI software with nsjail. 
Nsjail was configured to use filesystem restrictions, chroot, and a seccomp-bpf filter to kill the process in case of a violation.
With these restrictive measures, the sandbox environment effectively neutralized the same exploit. This demonstrated nsjail's capability to mitigate some vulnerabilities.

\subsubsection{Limitations and Further Considerations}
Despite this success, nsjail only partially eliminates the danger if the underlying vulnerability within the SUCHAI software remains unpatched. The possibility of more sophisticated exploits capable of circumventing the current sandbox's restrictions remains a proper concern. Other possible approaches could have been to use multiple nsjail instances to isolate individual command execution tasks rather than the entire software framework or running multiple SUCHAI FS instances, separating groups of tasks based on the shared libraries and funtionalities.

\subsection{Experiment with the SALSAT IPU Software}

The Spectrum AnaLysis SATellite (SALSAT)~\cite{salsat_mission} mission aimed to deploy a nanosatellite dedicated to spectrum analysis. Its primary payload, the Spectrum Analysis of LEO Satellite Allocations (SALSA) board, facilitated the monitoring of the VHF and UHF amateur radio bands, as well as scientific bands within the S band. Successfully launched on September 28, 2020, SALSAT entered a sun-synchronous orbit at an altitude of approximately 575 km and has been operational continuously since then \cite{freymuth22}. \newline
SALSAT feature redundant onboard computers (OBCs) and updated Payload Data Handling Units (PDHs). The PDH played a crucial role in managing the primary payload SALSA and various secondary payloads, including a camera, a Fluid Dynamic Actuator (FDA), and the i.MX-7 Linux on a Chip System ~\cite{phytec}, serving as an Image Processing Unit (IPU) \cite{vu19}. \newline
Using a Linux system enabled quick access to powerful image processing tools such as OpenCV, ImageMagick, and various Python libraries. 
To avoid having to implement all routines in one applications running multiple threads, the adoption of middleware became essential. As a result the ROS 1 middleware was integrated into the satellite. A diagram showing the architecture is available in the Appendix~\ref{sec:appendix:SALSAT_diagram}.

\subsubsection{Experimentation Setup and Methodology}
The experimentation with the SALSAT IPU Software necessitated a specific setup, primarily due to the software's requirement to run on Ubuntu 16.04. This older version of Ubuntu supports only Linux Kernel version 4.15, which conflicts with the requirements of the latest version of nsjail, tailored for Linux Kernel 5.4. Consequently, an older version of nsjail, version 3.2, released in 2022, was employed for compatibility.
The process of configuring nsjail to strike a balance between robust security measures and maintaining the full functionality of the SALSAT features presented further challenges. This balancing act was particularly complex due to the operational dynamics of ROS 1, which involves intricate communication between multiple executables and services. Unlike its successor, ROS 2, ROS 1's configuration is less elastic, necessitating careful tuning of both ROS 1 and nsjail settings to ensure smooth operation.
The optimal configuration was achieved by running only the command parser and execution node of ROS 1 within the nsjail sandbox, both responsible for received commands from the ground and executing them on the IPU. This approach effectively secured the most exposed surface area of the software. The final nsjail setup extensively employed read-only bind mounts, allowing ROS 1's access to its necessary libraries, chroot, and a seccomp-bpf filter.
\subsubsection{Vulnerability Introduction and Impact}
Mirroring the approach taken in the SUCHAI software experiment in Section \ref{suchai_vulnerability} a similar vulnerability was intentionally introduced into the SALSAT IPU Software. This command injection vulnerability was incorporated within the command parsing and execution ROS node, replicating the same type of security flaw previously implanted. The vulnerability exploitation could lead to executing of unauthorized commands via the ROS shell, representing a simulated yet realistic security threat.
In the absence of a sandbox, this vulnerability in the SALSAT IPU Software was, again, exploitable, posing a threat to the entire mission.

\subsubsection{Effectiveness of Sandbox Implementation}
As we saw with the SUCHAI experiment, the previously exploitable vulnerability was effectively neutralized upon running the ROS node with the nsjail sandbox using a similar configuration as the SUCHAI experiment. Despite the stringent constraints imposed by nsjail on system calls and file system access, the SALSAT software continued to function optimally.

\subsubsection{Limitations and Further Considerations}
The implementation of nsjail, as observed in the SUCHAI experiment, serves primarily as a mitigation tool rather than a complete elimination of vulnerabilities. A sophisticated attacker might theoretically exploit some of the syscalls required to run ROS (therefore not blocked by the seccomp-bpf filter), devising new exploits capable of bypassing nsjail's restrictions. Additionally, configuring nsjail to fully allow ROS functionalities proved time-intensive for a single node, raising concerns about scalability with multiple nodes.

\subsection{Integration to our current mission}
Following the encouraging outcomes of the experiments, the focus shifted to exploring the integration of nsjail with the new satellite currently under development. As mentioned before the satellite is based on Linux and utilizes ROS 2 as its middleware and Rust as the primary programming language.
Recognizing the similarities yet distinct differences between ROS 1 and ROS 2, the experiment aimed to assess the compatibility of ROS 2 with nsjail.
The testing involved running multiple ROS 2 nodes in separate nsjail environments, and it was conducted using simple and basic nodes. These tests verified the compatibility with all the ROS2 basic features, such as services, topics, and actions, on top of which all the other apps are built. The ability of these nodes to communicate effectively within nsjail environments suggested suitability for use in this scenario.
A notable observation was the relative ease in constraining ROS 2 executables compared to the earlier experience with ROS 1 in the SALSAT framework. This ease of configuration could be attributed to the inherent differences between ROS 1 and ROS 2, or it might be due to the simpler nature of the tested nodes, lacking extensive dependencies and requirements. The requirement for fewer parameters and bind mounts in ROS 2 could suggest the potential for more streamlined sandboxing processes in future implementations. 

However, a limitation arises when considering the integration of nsjail within the our new satellite framework. The current mode of operation for nsjail, primarily through terminal commands, is not feasible for real satellite applications. Satellite operations typically involve command scheduling via packets, and direct access through SSH to a shell is impractical.
To address this, two potential solutions are being considered. The first involves transforming nsjail into a library, allowing direct integration into the software. The second solution contemplates invoking the compiled nsjail binary from within the framework. However, both solutions are in preliminary stages and have yet to be implemented and tested.

Another aspect under discussion is what sections of the code should be contained inside a sandbox environment. One strategy is to run almost every ROS 2 application in its own nsjail sandbox, with configurations tailored to allow access only to necessary resources. This approach would involve modifying the ROS 2 launch system, a tool that automates running multiple nodes with a single command and is necessary in a complex scenario. By integrating nsjail into the ROS 2 launch process, certain groups of applications or nodes could be executed within a sandbox, replacing direct node calls. Alternatively, the focus could be on isolating only the command parser or handler nodes. This would involve spawning a new nsjail environment for each received command, limiting sandboxing to the most vulnerable code segments.

The decision on which strategy to adopt is pending. The reason is twofold: firstly, there is need for further testing with the actual framework and secondly, the design details of the satellite framework itself are still being finalized.

\section{Related work}

Some previous research discussed software isolation in space and embedded systems. Santangelo~\cite{quicksat2014} has shown how it is possible to achieve isolation on small satellites with the use of the Xen Space Hypervisor~\cite{xen_hypervisor}. Similarily Xtratum~\cite{xtratum} has demonstrated the application of hypervisor in space.
However, without hardware support hypervisors generally cause a severe performance overhead~\cite{djordjevic2021file}, and therefore this approach was not suitable for our mission. 

Additionally, B\"{a}ckman et al.~\cite{AppSecurityForEmbeddedSys} investigated application sandboxing in embedded systems operating on Linux. Their research primarily focused on determining the compatibility of various sandboxing technologies with the ARMv7 processor, as well as assessing whether their performance overhead was within acceptable limits.

\section{Conclusion}
Our study demonstrates the implementation of application sandboxing in small satellites. We established criteria for selecting a suitable sandboxing mechanism and validated this approach through practical application on two in-orbit satellites: Suchai and SALSAT. The sandboxing mechanism effectively isolated critical applications, mitigating the risk of complete satellite control in the event of a security breach. This approach not only proved viable in our satellite payload development but also showed adaptability to existing satellite frameworks. Our findings provide a foundation for future advancements in small satellite security, emphasizing the importance of application sandboxing measures in the evolving space technology landscape.

\section*{Acknowledgments}
We thank our anonymous reviewers for their helpful suggestions towards improving this paper. We further thank the Saarbrücken Graduate School of Computer Science for their funding and support.




\bibliographystyle{IEEEtranS}
\bibliography{references.bib}

\appendix
\section{Appendices}\label{sec:appendix}

\subsection{SALSAT IPU Architecture}\label{sec:appendix:SALSAT_diagram}
\begin{figure}[htbp]
  \centering
  \includegraphics[width=9cm]{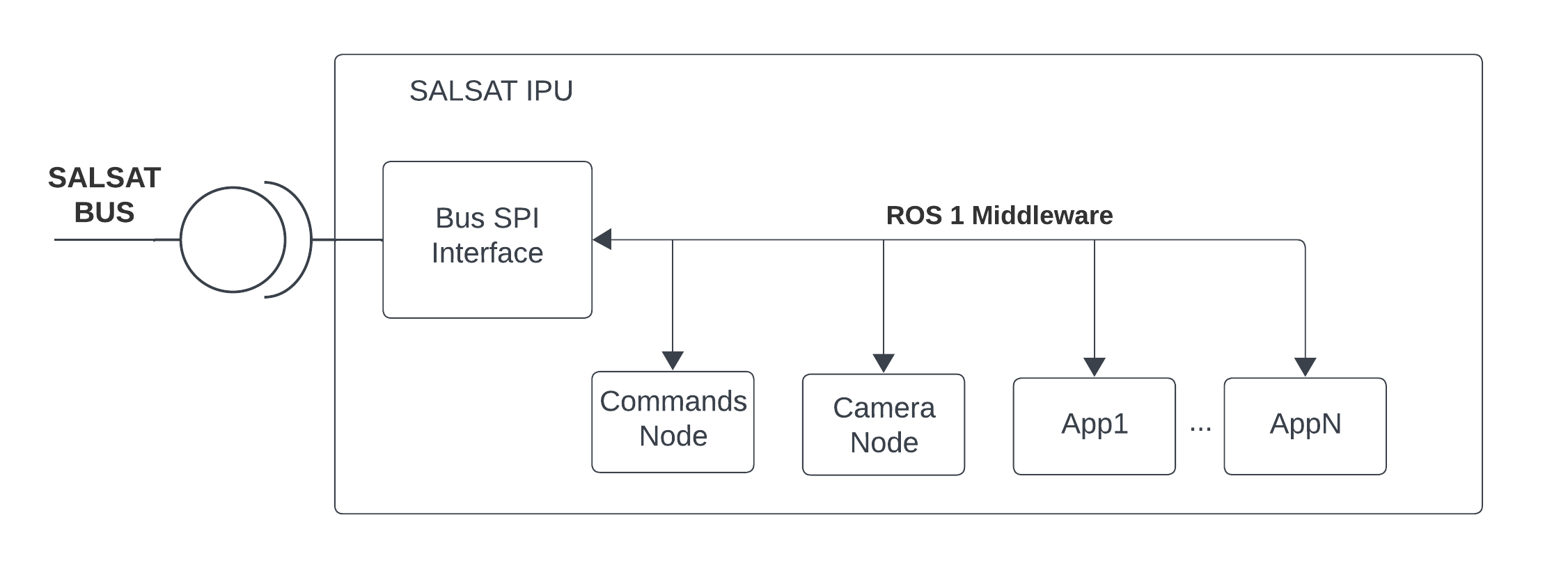}
  \caption{Diagram of the SALSAT IPU Architecture}
  \label{fig:salsat_ipu_architecture}
\end{figure}

The Figure~\ref{fig:salsat_ipu_architecture} shows the architecture of the SALSAT IPU software. It relies on ROS 1 middleware to enable communication between different nodes: BUS SPI Interface, the Commands Node, the Camera Node, and other apps.
The SPI bus allows the nodes to communicate with other satellite components or the ground.

\subsection{SUCHAI Architecture}\label{sec:appendix:SUCHAI_diagram}
\begin{figure}[htbp]
  \centering
  \includegraphics[width=9cm]{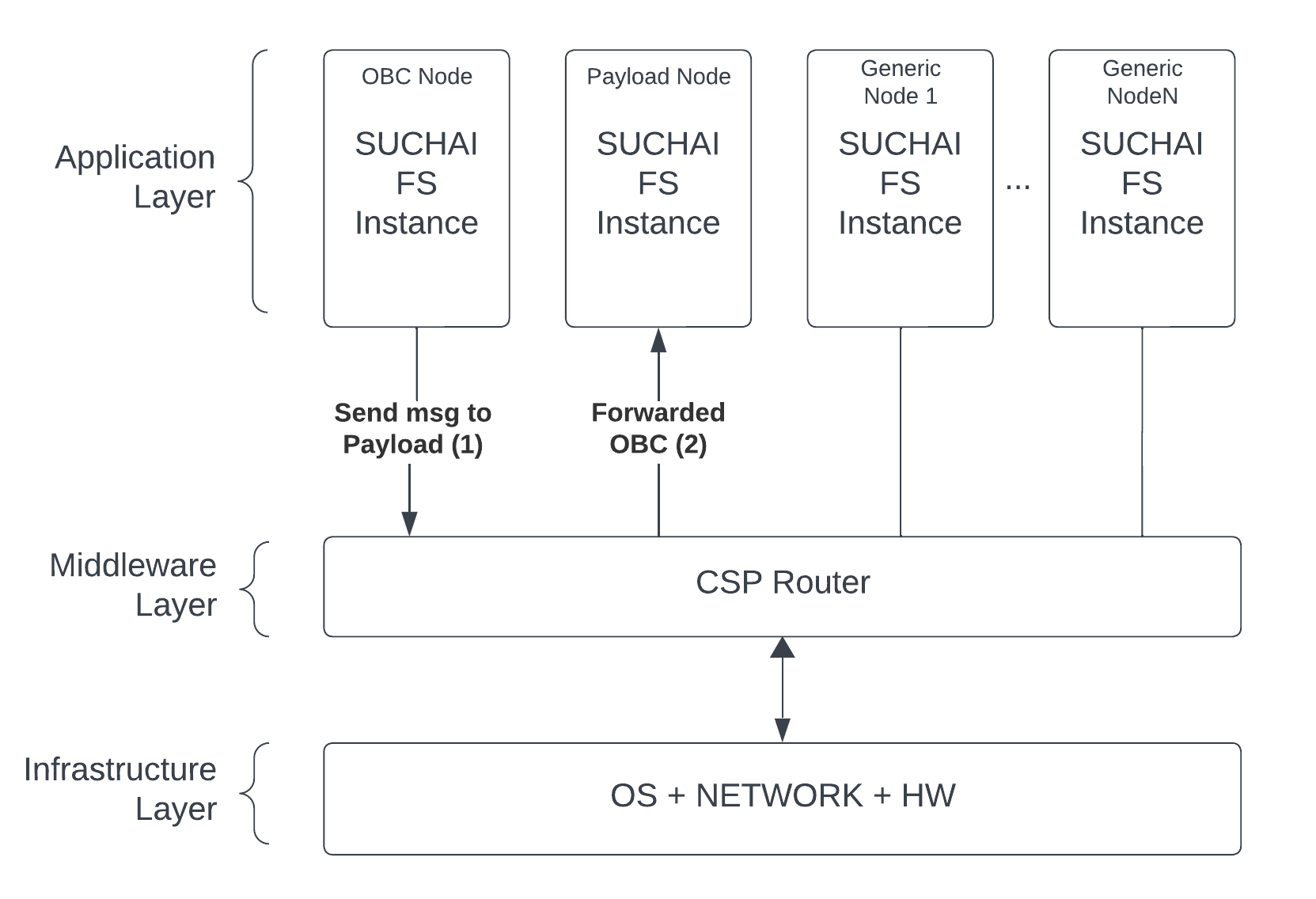}
  \caption{Diagram of the SUCHAI Software Architecture}
  \label{fig:suchai_architecture}
\end{figure}

The Figure~\ref{fig:suchai_architecture} shows the architecture of the SUCHAI software. It relies on the CSP Router and CSP protocol to enable communication between different nodes, for example, the instance running on the OBC and the one running the Payload. 
In the example, the OBC node sends a message to the Payload node through the CSP router that acts as a middleware-like layer.

\subsection{Our Mission - RACCOON OS Architecture}\label{sec:appendix:RACCOON_diagram}
\begin{figure}[htbp]
  \centering
  \includegraphics[width=9cm]{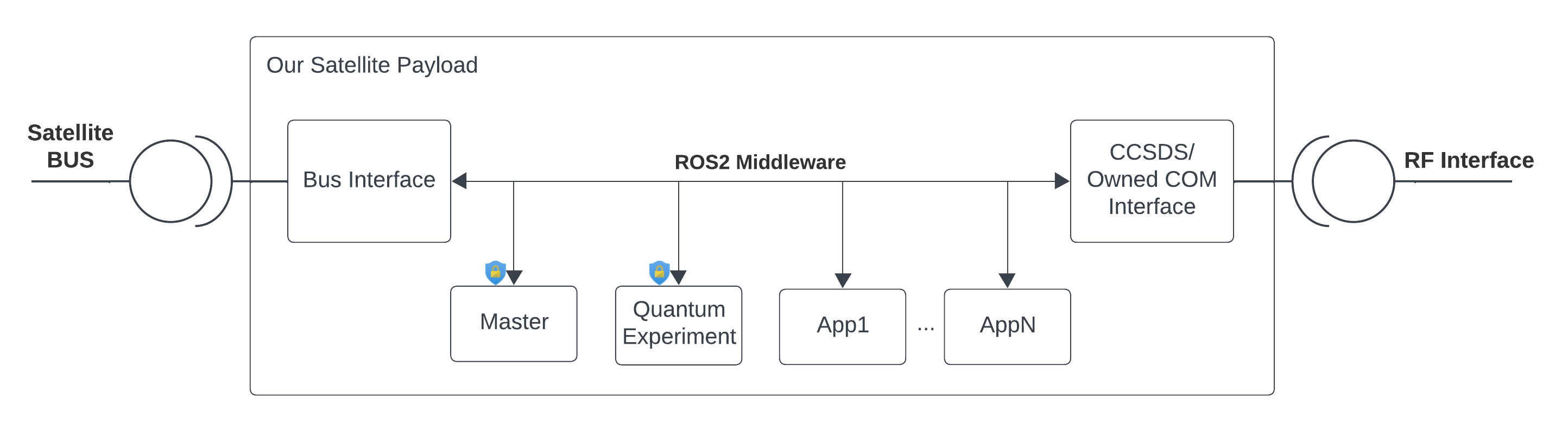}
  \caption{Diagram of the architecture planned for our satellite}
  \label{fig:raccoon_architecture}
\end{figure}

The Figure \ref{fig:raccoon_architecture} shows the architecture of our mission. The ROS 2 middleware will allow communication between the BUS (external satellite) and our onboard computer (payload) and between the different applications running on the system. As specified in Section \ref{sec:req:threat_model}, the Master and Quantum Key Distribution Algorithm nodes are the most valuable and must be isolated at all costs. The Quantum Key Distribution Algorithm is also called "Quantum Experiment" as in Figure \ref{fig:raccoon_architecture}.

\end{document}